\documentclass[aps,twocolumn]{revtex4}
\usepackage{graphicx,subfigure}
\usepackage{amsmath,amssymb}
\usepackage{bm}

\newcommand{\be}{\begin{eqnarray}}
\newcommand{\ee}{\end{eqnarray}}

\def\slashchar#1{\setbox0=\hbox{$#1$}           
   \dimen0=\wd0                                 
   \setbox1=\hbox{/} \dimen1=\wd1               
  \ifdim\dimen0>\dimen1                        
 \rlap{\hbox to \dimen0{\hfil/\hfil}}      
  #1                                        
 \else                                        
    \rlap{\hbox to \dimen1{\hfil$#1$\hfil}}   
    /                                         
 \fi}                                         %

\begin{document}

\title{Interacting Ensemble of the Instanton-dyons \\
and Deconfinement Phase Transition in the SU(2) Gauge Theory}

\author{ Rasmus Larsen   and Edward  Shuryak }

\affiliation{Department of Physics and Astronomy, Stony Brook University,
Stony Brook NY 11794-3800, USA}

\begin{abstract}
Instanton-dyons, also known as instanton-monopoles or instanton-quarks, are topological constituents of the instantons at nonzero temperature and holonomy.
We perform numerical simulations of the ensemble of interacting dyons for the SU(2) pure gauge theory.
Unlike previous studies, we focus on back reaction on the holonomy and the issue of confinement.
We calculate the free energy as a function of the holonomy and the dyon densities, using standard Metropolis Monte Carlo and integration over parameter methods.
We observe that as the temperature decreases and the  dyon density grows, its minimum indeed moves from small 
holonomy to the  value 
corresponding to confinement. We then report various parameters of the self-consistent 
ensembles as a function of temperature, and investigate the role of inter-particle correlations.
\end{abstract}
\maketitle

\section{Introduction}

QCD description of strongly interacting matter at finite temperature $T$ has originated from 1970's.
At first, its high temperature phase -- known as Quark-Gluon Plasma, QGP -- has been studied
using weak coupling methods, see e.g. reviews \cite{Shuryak:1980tp,Gross:1980br}.

 The interest then 
switched to non-perturbative phenomena, related with the topological solitons of various dimensionality
and two basic non-perturbative phenomena: confinement and chiral symmetry breaking.
 Instantons  \cite{Belavin:1975fg}, the Euclidean 4-dimensional topological solitons, have at high $T$ the sizes $\rho\sim 1/T$
and appear with the probability \be n_{instantons}\sim exp[-8\pi^2/g^2(T)]\sim \left({\Lambda \over T}\right)^b  \ee 
where the power is the one loop beta function coefficient, $b=11 N_c/3$ for $SU(N_c)$ gauge theory.  
So, at high $T$ the density is small and the topology is not important. Conversely, 
as  $T$
decreases, the instanton density grows rapidly, till they become an
 important ingredient of the
gauge fields in the QCD vacuum.
Chiral anomalies induce existence of the fermionic zero modes of instantons, which generate the so called 't Hooft effective interaction of $2N_f$ fermions, which 
 explicitly violates the $U_A(1)$ chiral symmetry. Furthermore, collectivization of the  zero modes
 create the so called Zero Mode Zone of quasi-zero eigenstates, which break spontaneously the $SU(N_f)$ chiral symmetry.
 Although those states includes only tiny  subset of all fermionic states in lattice numerical simulations,  they are 
the key elements of the  chiral symmetry breaking and the hadronic spectroscopy. The so called Interacting Instanton Liquid Model (IILM) has been developed, including  't Hooft interaction
 to all orders,  for a review  see
\cite{Schafer:1996wv}.

As the temperature decreases from the high-$T$ regime, another important phenomenon
is appearance of nontrivial expectation value of the Polyakov line.
For the SU(2) gauge theory we will be discussing in this work,
it is related to the so called holonomy parameter by $\langle P \rangle=cos(\pi\nu)$
 (for explicit notations see Appendix \ref{secHolonomy}). While at high $T$ it vanishes $\nu\rightarrow 0$,
at temperatures at and below the critical value $T_c$ it reaches the so called ``confining value" $\nu=1/2$
at which the Polyakov line vanishes. This leads to switching out quark and gluon degrees of freedom, and
transition from QGP to hadronic matter.
Study of the instantons at  nonzero holonomy has lead  Lee,Lu,van Baal and Kraan \cite{Lee:1998bb,Kraan:1998sn}
to the so called KvBLL caloron solution, which revealed  that at $\nu\neq 0$ the instantons get split into $N_c$
(number of colors) (anti)dyons, (anti)self-dual 3d solitons with nonzero (Euclidean) electric and magnetic charges. 
(Details are in Appendix \ref{secDyons}). Because of long-range forces
between these dyons, we will thus refer to instanton-dyon ensemble as the ``dyonic plasma".

Unlike instantons, the instanton-dyons interact directly with the holonomy. 
Diakonov \cite{Diakonov:2009ln}  suggested that back reaction of the 
dyon free energy on holonomy is responsible for confinement phase transition
 but was unable to show it.   Poppitz, Schaefer and Unsal \cite{Poppitz:2012nz}
  had shown that instanon-dyon confinement does occur  in 
a very specific ``controlled setting", a 
supersymmetric theory compactified on  $R^3\otimes S^1$ with a  small spatial circle and periodic fermions.
The smallness of the circle, like high $T$, 
 makes the coupling weak. The periodic fermions preserve supersymmetry and  cancel the perturbative holonomy 
 potential $V_{GPY}(\nu)$, which allows confinement to be induced even by  an exponentially 
small density of the dyons.
These authors have been able to trace the crucial effect to the repulsive dyon-antidyon interaction
inside the dyon-antidyon pairs, which they call ``bions".

 A phenomenological model showing that repulsive interaction between them, modeled by an excluded volume, 
  has been proposed for QCD-like theories by Shuryak and Sulejmanpasic \cite{Shuryak:2013tka}, which reached qualitative description of the deconfinement phase transition and 
 other properties of the thermal $SU(2)$ pure gauge system above $T_c$, in qualitative agreement
 with available lattice data. We will discuss similar model in section \ref{sec_excluded}, before we embark on numerical simulations.
 
 Although the interaction between the instanton-dyons have been studied for a long time, the leading-order
 effect -- classical dyon-antidyon interaction has been missing. The corresponding studies, deriving the
 so called ``streamline" set of configurations via the gradient flow
 method, has been done in our previous work  \cite{Larsen:2014yya}.
 This classical interaction turns out to be weak in relative terms $\delta S \ll S$ but, still being classical $\delta S\sim 1/g^2$,
 rather large numerically $\delta S=O(1)$ to induce significant correlations in the ensemble.
 
 Liu, Shuryak and Zahed \cite{Liu:2015ufa} 
have recently shown that one can incorporate this classical interaction by the  mean field
techniques, but only if the ensemble is dense enough to generate sufficient screening. In
terms of the temperature, this treatment applies only for $T<T_c$.

 The goal of our present work is to study the instanton-dyon ensemble by the direct Monte-Carlo
 simulation, without the {\em mean field} or any other approximations. As we will show, 
 this will allow us to cover both a  dilute and a dense regimes, and see in details the
 transformation of the holonomy potential which drives the deconfinement transition.
 In a way, this work is complementary to \cite{Liu:2015ufa}, since our main focus is
 at temperatures $T \geq T_c$.
 
 Technically, the details of the setting to a large extent follow the first Monte-Carlo simulations  
 of the instanton-dyon plasma  by Faccioli and Shuryak \cite{Faccioli:2013ja}.
 One major difference is the inclusion of the classical ``streamline" interaction which were not known
 at the time of that work. The other is that that paper focused on the role of fermions and  chiral symmetry breaking
 rather than confinement.  (We expect to report on our next paper, with fermions, soon.)

The paper is structured as follows: 
  standard information about our notations, the holonomy potential and the instanton-dyons are
  delegated to sections of the Appendix.
 We start in section \ref{sec_excluded} by introducing a simple model which illustrates the main physics under discussion.
 Then in section \ref{sec_potentials} we explain the setting of the simulations,  the  interaction between the
 instanton-dyons  and the moduli spaces which provides the measure in the partition function. 
 In section \ref{sec_simulations} we describe how we make the actual simulations and evaluate the free energy.
 The back reaction of the ensemble on the holonomy potential is described in section \ref{sec_reaction},
 which is followed by ``self-consistency" study of the parameters in the section \ref{sec_self}. The physical results
 are summarized in the section \ref{sec_results}: those include the holonomy potential and the screening masses,
 as well as the densities of all types of dyons. 
 
 \section{An excluded volume model} \label{sec_excluded}
 To understand the main physics involved and the qualitative behavior of the ensemble, including the confinement phase transition,
 we start by a discussion
 of a simplified model in which the only interaction
 is the repulsive core, making 
   the volume occupied by each particle unavailable to others. 
   It is similar in spirit to that proposed by Shuryak and Sulejmanpasic \cite{Shuryak:2013tka},
   but is somewhat closer technically to the simulations to follow.
   
We work in dimensionless units and we therefore define the 3-volume $\tilde V_3=T^3 V_3$, the density $n_i=\frac{N_i}{\tilde V_3}$, and the free energy density as $\frac{ F}{T \tilde V_3 } = \ f$.   More information on units and notations 
 can be found in Appendix.  
   
   The effect of the excluded volume is accounted for in a very schematic way, by
   cutting off the partition function when the amount of available volume vanishes. The volume of the M and L dyons scale by $1/\nu^3$ and $1/\bar{\nu}^3$ respectively, with $\bar{\nu}=1-\nu$. 
 We thus have the partition function in dimensionless units
 as a sum limited from above
\begin{eqnarray}
&Z& = \sum_{M,L}^{\tilde V_3/(V_0)<M/\nu^3+L/\bar{\nu}^3} \exp\left(-\tilde V_3\frac{4 \pi^2}{3}\nu ^2 \bar{\nu}^2\right) \\
&\times & \left[\frac{1}{M!L!}(\tilde V_3 d_\nu)^M(\tilde V_3 d_{\bar{\nu}})^L \right]^2   \nonumber \\
d_\nu &=&\Lambda \nu ^{8\nu /3} S^2\exp(-S\nu) \\
S &=& \frac{8\pi^2}{g^2}
\end{eqnarray}
 If the upper limit is ignored, and the volume goes to infinity,  $log(Z)/\tilde V_3\rightarrow -\frac{4 \pi^2}{3}\nu ^2 \bar{\nu}^2 + 2(d_\nu+d_{\bar\nu}) $, the perturbative homonomy potential plus the contribution of the noninteracting dyons.
 This explains that $d_\nu$  -- the semiclassical dyon amplitude --  is their density if all interactions are ignored. The parameter $S$ is in fact the classical action 
 of the caloron, or $L+M$ system. The square comes from assuming the same amount of dyons and antidyons.

 If we are in the confining phase, $\nu=\bar\nu=1/2$, all dyons have the same sizes, and it is easy to introduce
 the excluded volume, for $N$ dyons via $$\tilde V_3^N\rightarrow \tilde V_3\left(\tilde V_3-V_{excluded})... (\tilde V_3-(N-1)V_{excluded} \right)
 $$
 However, since in general we have $L,M$ dyons of different sizes, the analogous expression becomes cumbersome.
 Experimenting with those, 
  we observe that similar results are obtained by simply cutting out the sum
at the number when there is no volume left, $\tilde V_3<V_0(M/\nu ^3+L/\bar{\nu}^3)$.
 where $V_0$ is the excluded volume normalized for a dyon at $\nu =1$. 
 
 Using Sterlings formula $n! \approx \sqrt{2\pi n}\left(\frac{n}{e}\right)^n$ for a large volume, we rewrite the sum as
\begin{eqnarray}
&Z& = \sum _{M,L}^{\tilde V_3/(V_0)<M/\nu^3+L/\bar{\nu}^3} \exp \bigg[ -\tilde V_3\bigg(\frac{4 \pi^2}{3}\nu ^2 \bar{\nu}^2 \nonumber\\
& & -2n_M\ln\left[\frac{d_\nu e }{n_M}\right]-2n_L\ln\left[\frac{d_{\bar{\nu}} e }{n_L}\right]\bigg)\bigg] \label{ModelZ}
\end{eqnarray}

The free energy given by $F(\nu)=-T log Z$ depends on $\nu$, located in the cutoff, in the dyon parameter $d_\nu$, and in the
 last perturbative term. This last term, if dominant, would select $\nu=0$ or $\bar{\nu}=0$.
 (The value of $S$, the action for a dyon with $\nu=1$, is treated as an input parameter.) 
If one wants the two former (dyonic) exponents to be dominant instead, and select some nontrivial $\nu$
the corresponding densities should be large enough. In the simplest
  confining case $\nu= \bar{\nu}=1/2$ and $d_\nu=d_{\bar{\nu}}$, one finds that the density has to be of the order of $1/4$ of the perturbative energy  to overpower it. 
 
\begin{figure}[t!]
  \begin{center}
  \includegraphics[width=7cm]{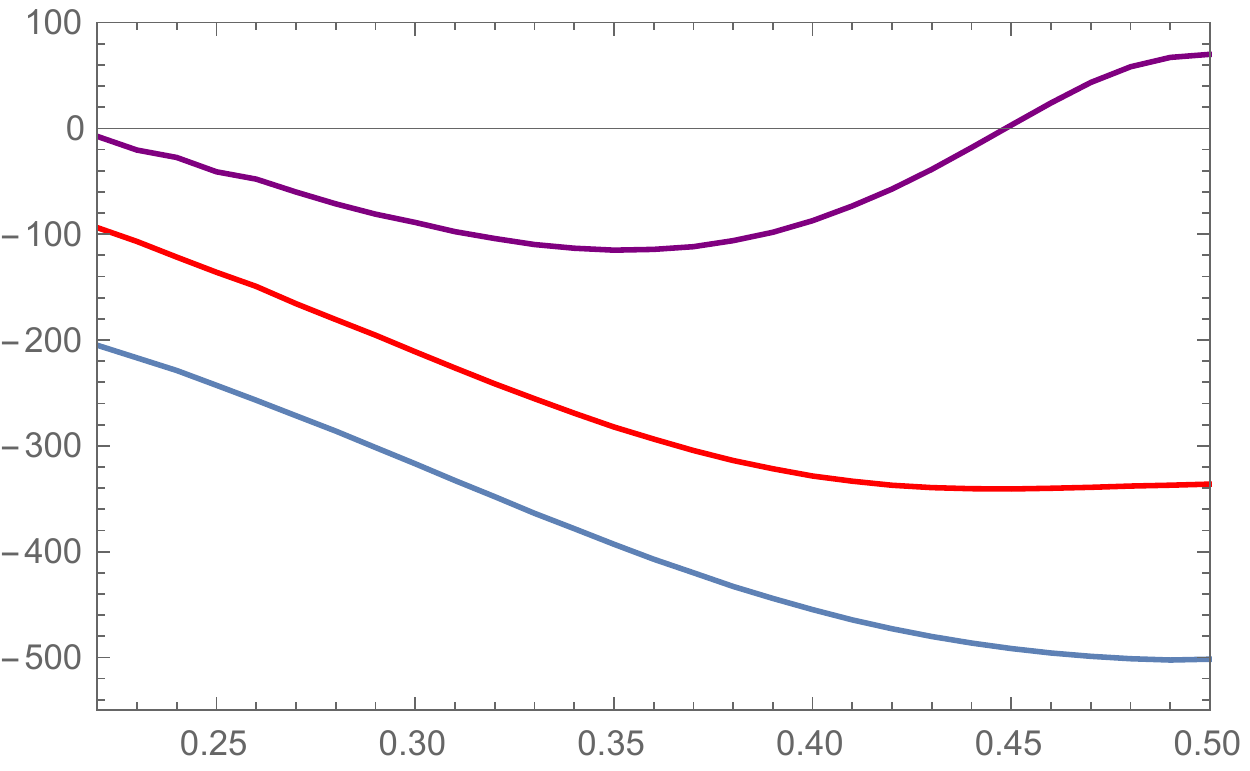}
    \put(-75,-5){$\nu$}
    \put(-210,60){$f$}
   \caption{ (Color online). Plot of the free energy density $f$ as function of holonomy $\nu$ for $\Lambda=0.5$ and $V_0=0.3$ at $g=4,3.5,3$, bottom to top. It is seen how the maximum as a function of $g$ goes further and further towards the confining value of $1/2$ as g goes up, and $S$ and $T$ go down.}
  \label{3dplot1}
  \end{center}
\end{figure}

The expression  (\ref{ModelZ}) is put  into Mathematica and the maximum is found, for
large enough volume, say  $V=900$. One finds a sharp peak in $N$ distribution, defining the density.  
 Finding the maximum as a we vary $\nu$, we get 
  $f(\nu)=-log Z/ \tilde V_3$  plotted in   
 Fig. \ref{3dplot1} for $g=4,3.5,3$. At  smaller g (larger S and higher $T$) the dyons are more suppressed and the free energy density $f$ has a minimum
 at smaller $\nu$.  For increasing coupling $g$ (decreasing  $S$ and  $T$), the minimum shifts 
 from zero, eventually to its confining value
 $\nu=1/2$.
 
\begin{figure}[t!]
  \begin{center}
  \includegraphics[width=7cm]{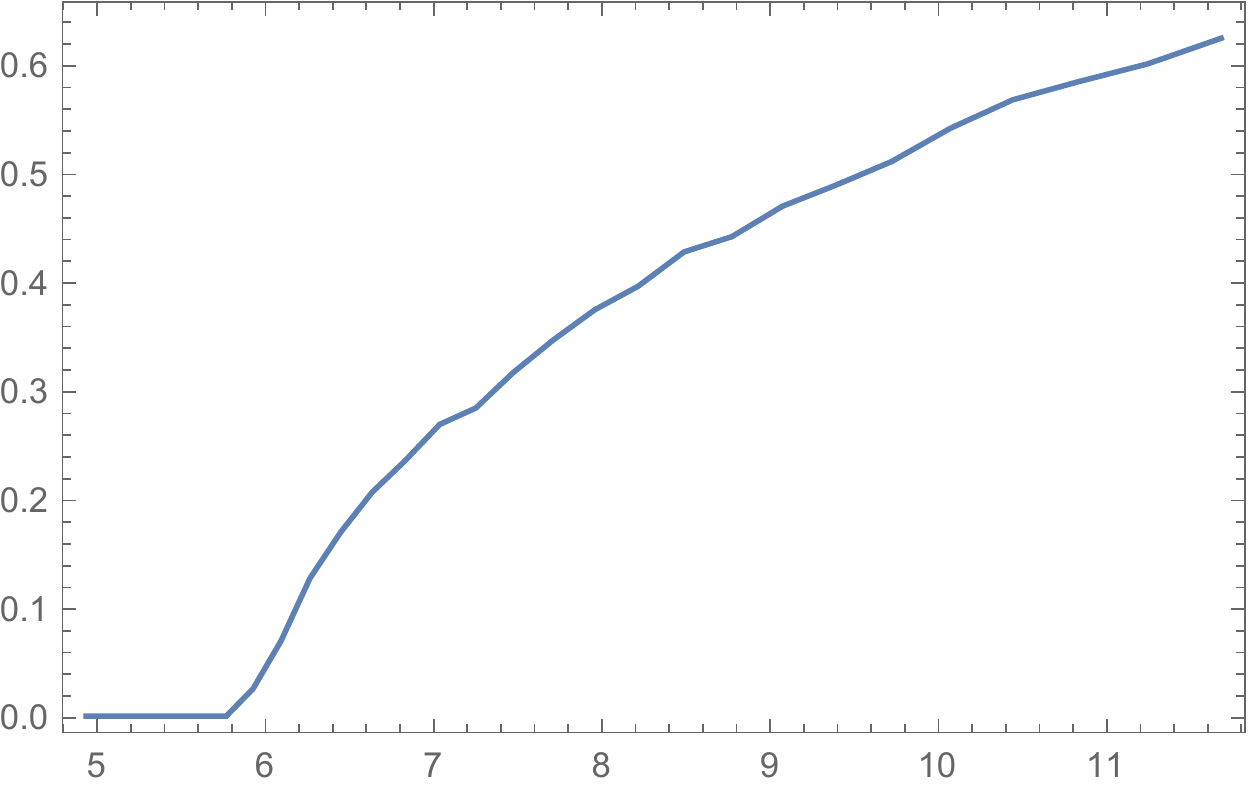}
 	 \put(-95,-5){$S$}
    \put(-210,60){$P$}
   \caption{ (Color online). Polyakov loop $P$ as a function of action parameter $S$ for $\Lambda=0.5$ and $V_0=0.3$. }
  \label{PL1}
  \end{center}
\end{figure}

 Instead of showing $\nu$ itself, one can also plot the average Polyakov loop, depending on it as
 $<P>=cos( \pi \nu)$, versus $S$ in  Fig. \ref{PL1}.
 The parameter $S$ grows with $T$ (see details in \ref{secHolonomy}). 
 In this model it is seen how the Polyakov loop continuously goes to $0$ at $S$ slightly smaller than 6. 
 In order to get a better perspective on how this happens, we  show the free energy versus
 the holonomy, as three curves in Fig \ref{3dplot1}. Note that in the critical case, the middle curve,
 the free energy $f(\nu)$ becomes very flat.

\begin{figure}[t!]
  \begin{center}
  \includegraphics[width=7cm]{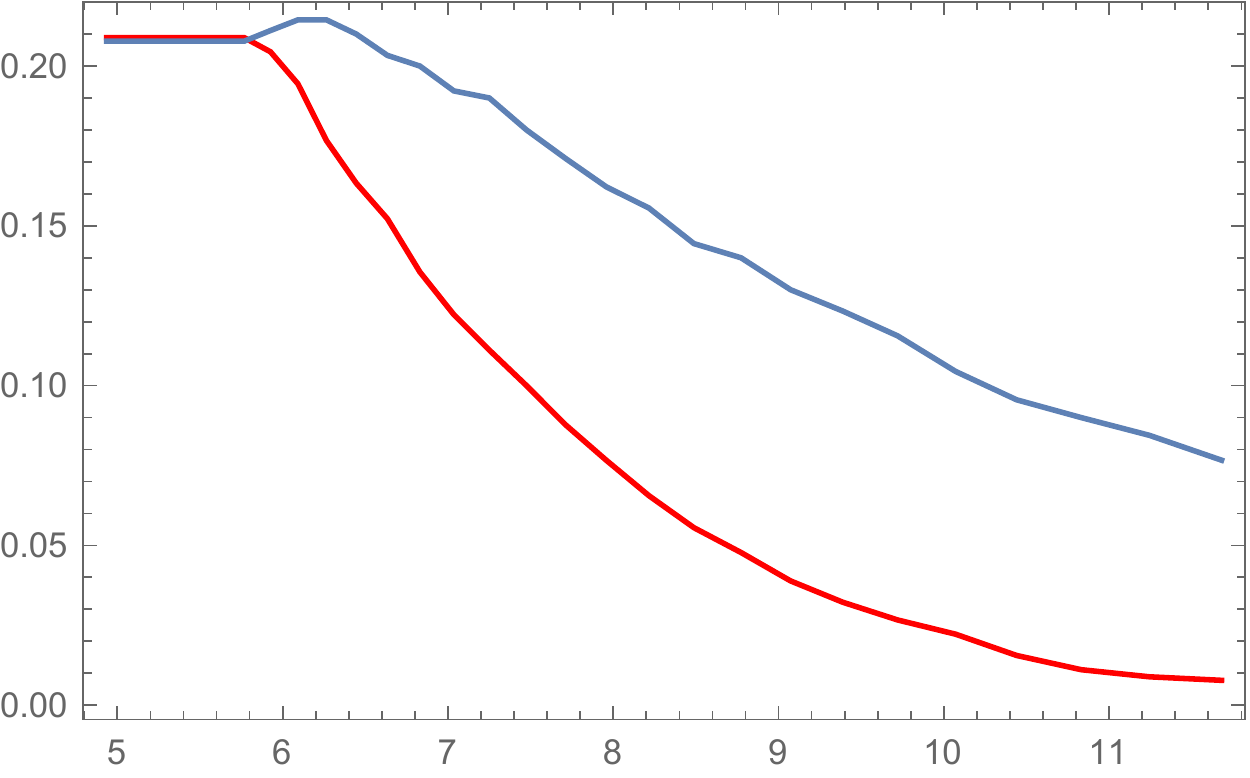}
  \put(-95,-5){$S$}
    \put(-210,60){$n$}
   \caption{ (Color online). Densities $n_i$ of $i=M$ or $i=L$ dyons as a function of the action parameter $S$, for $\Lambda=0.5$ and $V_0=0.3$. Note that the two densities are different at $S>6$.}
  \label{Density1}
  \end{center}
\end{figure}

\begin{figure}[t!]
  \begin{center}
  \includegraphics[width=7cm]{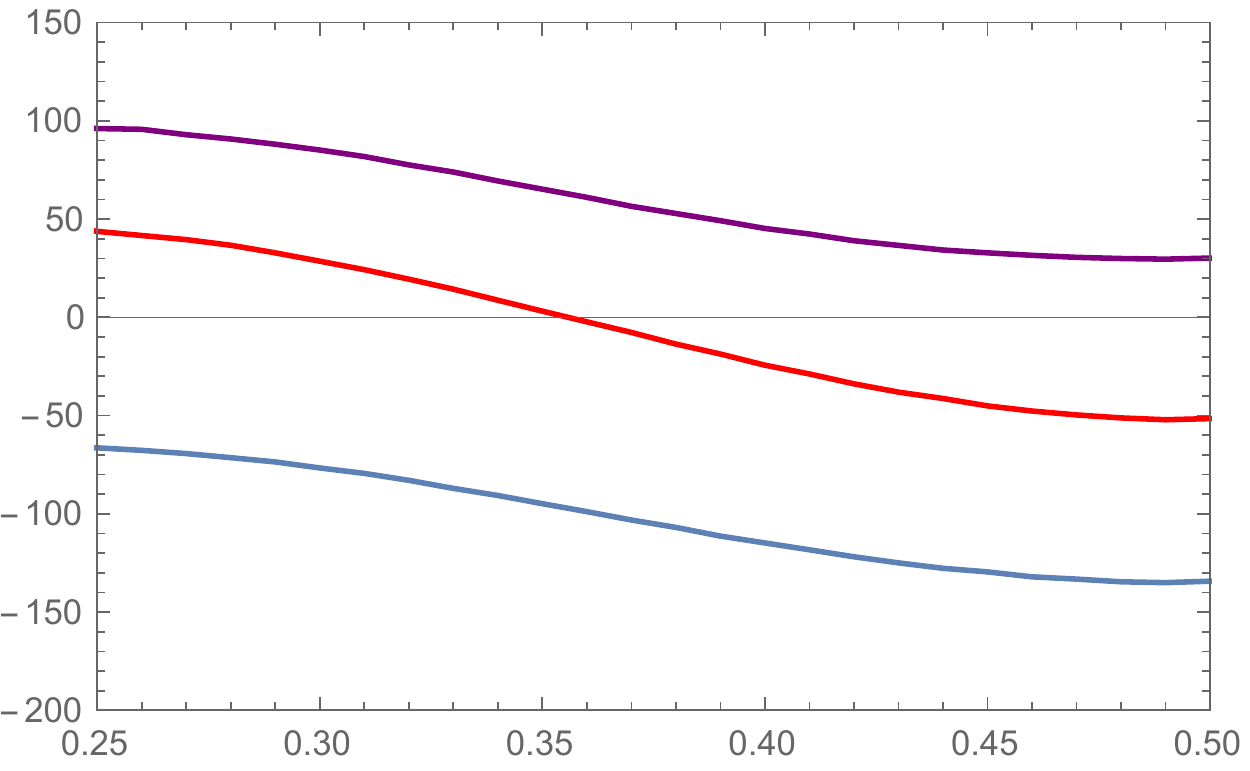}
  \put(-95,-5){$\nu$}
    \put(-210,60){$f$}
   \caption{ (Color online).
   The same as Fig.\ref{3dplot1}, but for larger excluded volume  $V_0=0.6$ (and slightly different $g=4,3.5,3.25$).
 Since $f=0$ corresponds to no dyons, a jump will be observed from confining holonomy $\nu=1/2$ to trivial holonomy $\nu=0$.}
  \label{3D2}
  \end{center}
\end{figure}

 As the repulsion (excluded volume) grows, the free energy increases, and start coming into the positive domain, see
Fig.\ref{3D2}. The $f>0$ formally means that the vacuum without any dyons, with  $f=0$, is lower than with them: 
too repulsive dyons drive themselves out of existence.
Obviously such situation is not physically possible.  

We then  show in Fig. \ref{Density1}, at above the deconfinement transition  the densities of 
different type ($M$ and $L$) dyons is different. We will see similar plots as a result of numerical simulations below.
Here let us only comment that direct evidences for $n_M>n_L$ in the deconfined phase
have been found on the lattice, and are related to the issue of chiral symmetry breaking for
fermions with different periodicity angle, see discussion in \cite{Shuryak:2012aa}.

 \section{The instanton-dyon interactions}\label{sec_potentials}
 Important new element is the inclusion of the {\em leading order} dyon-antidyon interaction 
 recently studied in our previous paper \cite{Larsen:2014yya}. We will use a slightly different parameterization that follows from the same data
\be 
\Delta S_{D\bar{D}} &=&  -2\frac{8\pi^2 \nu}{g^2}(\frac{1}{x}-1.632e^{-0.704x}) \nonumber \\
x &=& 2\pi \nu r T
\ee
for distances larger than $x>4$. 

It has been found that at interdyon distance $x=4$ the streamline terminates by rapid annihilation of the magnetic charges. If dyons are put at smaller distances, they repel till distance 4, before annihilation. Those configurations were
not yet studied in detail, and thus our potential
for $x <4$  is a reasonable guess. We cut the potential off at a distance $x_0=r_0T(2\pi \nu)$ with a core which we describe by
\begin{eqnarray}
\Delta S_{D\bar{D}} = \frac{\nu V_0}{1+ \exp\left[\sigma(x-x_0)\right]}
\end{eqnarray}
where we scale the core also by $\nu$ since we want the interaction to disappear at $\nu=0$.

It is perhaps worth reminding why the attractive Coulomb-like interaction is there. Selfdual and antiselfdual
objects have electric and magnetic forces canceling each other, but the effect still comes 
via $A_4$ and 
 the non-linearity of the field strength tensor. It is well known from the analytic results for calorons 
 that such interaction  can be described purely from the charges in table \ref{tab1} such that the interaction is $V = (e_1e_2+m_1m_2-2h_1h_2)\frac{4\pi}{g^2}\frac{1}{r}$. Note
 that this interaction is repulsive for $M\bar{L},L\bar{M}$ channels, a result that has been checked 
 by us (after submission of our previous paper \cite{Larsen:2014yya}).
 Note also that this expression also agrees with
  zero classical interaction between sectors that are completely self-dual ($LM$) or anti-self-dual.


 The volume element of the space of collective variables we will use is in the form of the so called Diakonov determinant
 \be  \sqrt{g}=det{G}
 \ee  
 \be G &=& \delta _{mn} \delta _{ij} ( 4\pi \nu_m-2\sum _{k\neq i}\frac{1}{T|x_{i,m}-x_{k,m}|} \\
 & & +2\sum _{k}\frac{1}{T|x_{i,m}-x_{k,p\neq m}|} ) \nonumber \\
 & & +2\delta_{mn}\frac{1}{T|x_{i,m}-x_{j,n}|}-2\delta_{m\neq n}\frac{1}{T|x_{i,m}-x_{j,n}|} \nonumber
 \ee
where $x_{i,m}$ denote the position of the i'th dyon of type m. The Diakonov determinant is a combination of the metric between a $M$ and $L$ dyon, which is true at any distance, and the metric between dyons of same type, which is only true at large distance, but does have the correct behavior of being repulsive for small separations. We therefore introduce a cutoff on the separation $r \to \sqrt{r^2+cutoff ^2}$, such that for one pair of dyons of same type, the diagonal goes to 0 for $\nu=0.5$, instead of minus infinity.

We use the same metric for the antidyons also.

It is observed that in case the density of $M$ and $L$ dyons are different, then in the infinite volume limit, the sum will diverge. We therefore regularize all the terms with $r \to r e^{M_D r T}$, where we work with the dimensionless Debye mass. 

The same exponentially damped effect due to the Debye mass is also applied to the Coulomb potentials. With this the interaction is given by 
\be
\Delta S_{D\bar{D}}  &=& \frac{8\pi^2 \nu }{g^2}\left( (e_1e_2-2h_1h_2)\frac{1}{x}+m_1m_2\frac{1}{x}\right)e^{-M_D r T} \nonumber \\
x &=& 2\pi \nu rT
\ee
for $r$ larger than  the core of size $r_0/(2\pi \nu)$ for all combinations except between dyons and their antidyon. For the dyon antidyon potential we have
\be 
\Delta S_{D\bar{D}} &=&  -2\frac{8\pi^2 \nu}{g^2}(\frac{1}{x}-1.632e^{-0.704x})e^{-M_D r T} \nonumber \\
x &=& 2\pi \nu r T
\ee
We include the core for both dyon antidyon interaction, but also for dyon dyon interaction due to the lack of a repulsion, which otherwise  destroys the simulation. We hope that such an interaction can be found due to loop calculations as it was done with instantons.
\be
\Delta S_{D\bar{D}}  &=& \frac{\nu V_0}{1+\exp\left[\sigma T(r-r_0)(2\pi \nu)\right]}
\ee

 \section{The setup} \label{sec_simulations}
 Like in \cite{Faccioli:2013ja}, instead of toroidal box with periodic boundary conditions in all coordinates, our
  simulations have been done on a $S^3$ sphere (in four dimensions),
  to simplify treatment of the long range Coulombic forces. In this pilot study we fix the total number of 
   dyons to 64. 
   We opt instead for multiple runs, displaying dependencies of the free energy on all parameters, sacrificing
   somewhat statistical accuracy. We do not use large computers, relying on multiple cores of standard GPU's
    of one standard computer.
   
   The radius of the sphere together with the ratio of M dyons to L dyons have been used to change their density. 
   
   Iteration of the system is defined as a loop in which each dyon has had its position changed  and the new action has then been accepted with the probability of $\exp(-\Delta S)$ via Metropolis algorithm. The typical number
   of iterations, for equilibration is 400 and after equilibration 1600.
  
In order to get the free energy we also use standard method. One can
differentiate with respect to an auxiliary parameter $\lambda$ introduced in front of the action and get
\begin{eqnarray}
e^{-F(\lambda)/T} &=& \int Dx \exp (-\lambda S(x))\\
\frac{\partial F}{\partial \lambda} &=& T\langle S\rangle
\end{eqnarray}
Since the free energy at $\lambda =0$ is known analytically, one can integrate up to get the free energy at $\lambda =1$.
When we do this we of course need to be careful about regions with a quick change in the action.



For the calculation of $\det G$ it has been 
observed by Bruckmann et al \cite{Bruckmann:2009nw} that it is only make sense if all eigenvalues are positive.
It was observed \cite{Bruckmann:2009nw} that,
 for randomly placed dyons this is typically not the case, unless density is very low. 
In \cite{Liu:2015ufa} this issue has been discussed further, with a conclusion that the Diakonov determinant
can remain positive definite at higher densities needed, but only with certain correlations in the dyon
locations enforced.

We have therefore used the Householder QR algorithm together with tridiagolization of the matrix $G$ \cite{Cosnuau:2014} to find the eigenvalues. 
We also redefine the potential as follows:

 If all eigenvalue are positive
\be
V_D &=& -\log [Det(G)] : V_D < V_{max} \\
V_D &=&  V_{max} : V_D > V_{max} 
\ee
and for one or more negative eigenvalues
\be
V_D &=&  V_{max}
\ee
This is done such that the excluded volume from the regions of negative eigenvalues are included, and at the same time to not create a region where the configuration is confined inside the region of negative eigenvalues. 
The possible great jump in energy from this definition means that we first integrate the free energy up to $\lambda =0.1$ finely with 10 points, and then with 9 points integrate from $\lambda =0.1$ to $\lambda =1$.

\section{The dyon back reaction: holonomy potential} \label{sec_reaction}
 In the absence of non-perturbative effects, there is only a perturbative
 interaction of thermal gluons with the holonomy, generating the perturbative Gross-Pisarski-Yaffe potential
  \cite{Gross:1980br} (\ref{GPY}) which disfavor confinement.  However, as  reproduced countless of times in any finite $T$
  lattice gauge simulations during the last 3 decades, the peak of the holonomy distribution
shifts to confinement at $T\approx T_c$.
 The corresponding effective potential
  $V(\nu)$ has been numerically studies and parameterized,  used in many models of finite-$T$
  QCD,  e.g. the so called Polyakov-Nambu-Jona-Lasinio (PNJL) model.
  Now our task is to derive such potential, stemming from back reaction of the instanton-dyons.

So, we
 add $V_{GPY}$
to the dyon free energy obtained from our simulations and determine the total free energy of the system
(obviously, assuming that there are no other relevant non-perturbative contributions). 
The dyon-induced partition function  is further split into two factors: 
one containing all factors which depend on parameters unchanged in a simulation sets, and the second one
related to dyons's collective variables.   
\be Z= Z_{unchanged} Z_{changed} \ee
The weight for one caloron ($L+M$ pair) was explicitly calculated in \cite{Diakonov:2004jn}:
at zero holonomy it agrees with the instanton result by 't Hooft. Part of the answer is 
the factor coming from metric volume element $\sqrt{g}$ in the space of $L,M$ collective variables.
 Later Diakonov \cite{Diakonov:2009ln} combined this result with previously known answer
 for metric of two monopoles of the same kind (e.g. $M,M$ pair) into an elegant expression for any number of $L,M$
 dyons now called Diakonov determinant 
  $\det G$.  Taking the dilute limit $r_{12}\rightarrow \infty $  in both cases both formulas reduce to the same $r_{12}$ dependence and
  one finds that the caloron weight from \cite{Diakonov:2004jn} needs to be divided by the factor $ (4\pi \nu)(4\pi \bar\nu)$ (see appendix \ref{secPartition})

\be Z_{unchanged} &=&\frac{ \Lambda^2}{(4\pi)^2} \left( \frac{8\pi ^2}{g^2} \right)^4 e^{-\frac{8\pi ^2}{g^2}} \nu ^{\frac{8\nu}{3}-1}\bar{\nu} ^{\frac{8\bar{\nu}}{3}-1}
\ee
Note that at the trivial holonomy $\nu\rightarrow 0$ limit $Z_{unchanged}$ is $\sim 1/\nu$: it is to be canceled by
the diagonal part of the $\det(G)$ in the second part.

Since we want to do the simulation for  different amount of $M$ and $L$ dyons we need to split up  the weight into a $M$ part and $L$ part and sum over all number of particles. We choose 
\be &&  Z_{unchanged} =  \sum _{N_M,N_L} \\
& &\left[\frac{1}{N_M!}\left(\Lambda \tilde V_3\left( \frac{8\pi ^2}{g^2} \right)^2 e^{-\frac{\nu 8\pi ^2}{g^2}} \nu ^{\frac{8\nu}{3}-1}/(4\pi)\right)^{N_M}\right]^2 \nonumber \\
&\times & \left[\frac{1}{N_L!}\left(\Lambda  \tilde V_3 \left( \frac{8\pi ^2}{g^2} \right)^2 e^{-\frac{\bar{\nu} 8\pi ^2}{g^2}} \bar{\nu} ^{\frac{8\bar{\nu}}{3}-1}/(4\pi)\right)^{N_L}\right]^2 \nonumber
\ee
where we use that the amount of dyons and antidyons is the same. We simplify this as 
\be Z_{unchanged} &=&  \sum _{N_M,N_L} \left[\frac{1}{N_M!}\left(\tilde V_3 d_\nu \right)^{N_M}\right]^2 \nonumber \\
&\times & \left[\frac{1}{N_L!}\left(\tilde V_3 d_{\bar{\nu}}\right)^{N_L}\right]^2 \nonumber \\
d_\nu &=& \Lambda \left( \frac{8\pi ^2}{g^2} \right)^2 e^{-\frac{\nu 8\pi ^2}{g^2}} \nu ^{\frac{8\nu}{3}-1}/(4\pi)
\ee

$ Z_{changed}$ is the interactions explained in section \ref{sec_potentials} and thus also depends on the number of particles
\be Z_{changed}&=& \frac{1}{\tilde V_3^{2(N_L+N_M)}}\int D ^3x \det (G) \exp( -\Delta D_{D\bar{D}} (x) ) \nonumber \\
\Delta f &\equiv & -\log (Z_{changed})\ee
where we have normalized such that $Z_{changed}=1$ for no interaction. Combining with the unchanged part and the purturbative potential we get in the limit $V\to \infty$ 
\be Z &=&  \sum _{N_M,N_L} \exp\bigg( -\tilde V_3\bigg[\frac{4 \pi^2}{3}\nu ^2 \bar{\nu}^2 -2 n_M\ln\left[\frac{d_\nu e }{n_M}\right] \nonumber \\
& &+2 n_L \ln\left[\frac{d_{\bar{\nu}} e  }{n_L}\right]+\Delta f\bigg]\bigg)
\ee

%
%

For $\tilde V_3\to \infty$ the partition function is completely dominated by the maximum of the exponent. Finding the free energy corresponds to finding the minimum of
\be
f &=& \frac{4 \pi^2}{3}\nu ^2 \bar{\nu}^2 -2n_M\ln\left[\frac{d_\nu e }{n_M}\right] \nonumber\\
& &-2n_L\ln\left[\frac{d_{\bar{\nu}} e }{n_L}\right]+\Delta f
\ee 

Note that as the dyon density increases, it changes its shape, producing
a non-trivial minimum at $\nu \neq 0$. Furthermore, at high density this minimum 
moves to $\nu=1/2$, the confining value.

The densities of both kinds of dyons
$n_L,n_M$ are not in general equal: the model should be able to do this by adding 
compensating charge to the whole sphere. In our model this is done by including the Debye mass.

%
%
%
%

\section{Self Consistency} \label{sec_self}

   The partition function we simulate depends on several parameters, 
    changed from one simulation set to another. Those include
   (i) the number of the dyons $N_M,N_L$; (ii) the radius of the $S^3$ sphere $r$;
 (iii)  the
   action parameter $S$;
  (iv) the value of the
   holonomy $\nu$, (v) the value of the
    Debye mass $M_D$; (vi) 
   the auxiliary factor $\lambda$, which is then integrated over as explained in section \ref{sec_simulations}.
 

\begin{figure}[t!]
  \begin{center}
  \includegraphics[width=8cm]{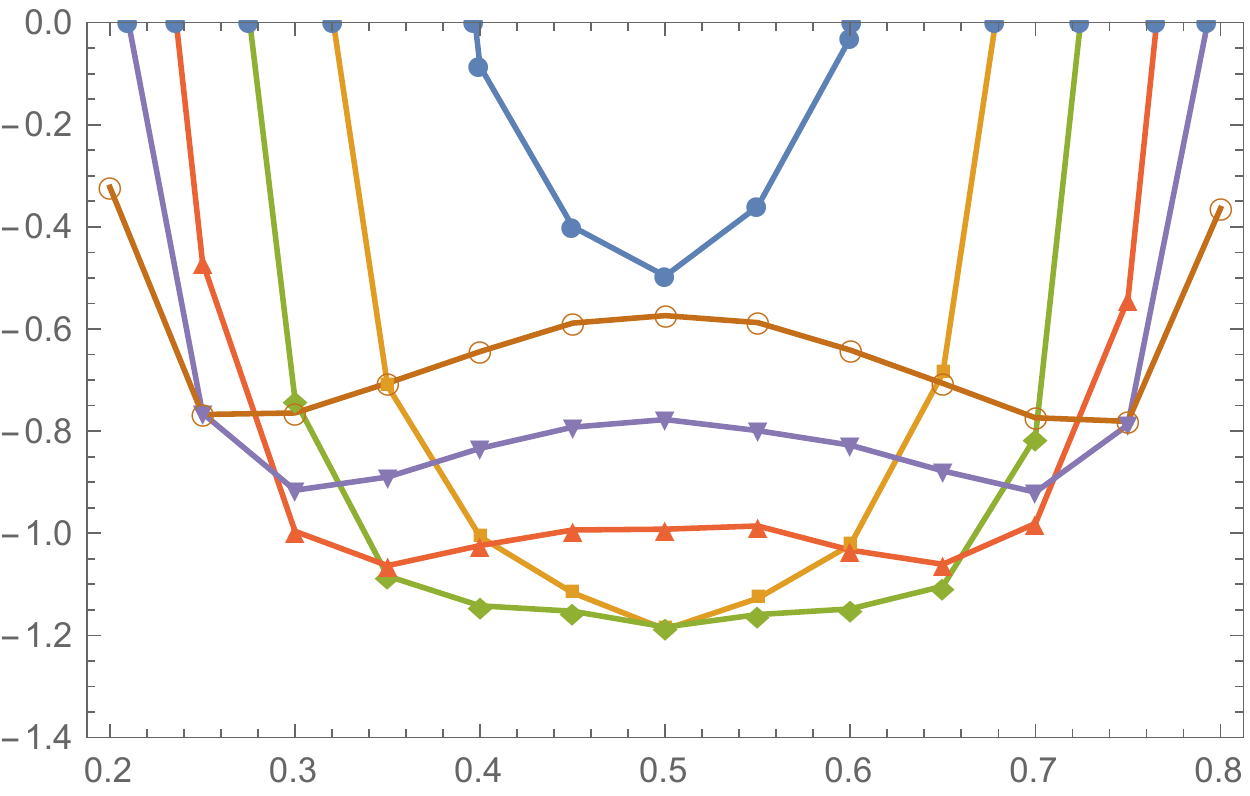}
  \put(-95,-5){$\nu$}
    \put(-230,70){$f$}
   \caption{ (Color online). Free Energy density $f$ as a function of $\nu$ at $S=6$, $M_D=2$ and $N_M=N_L=16$. The different curves corresponds to different densities. $\bullet$ $n=0.53$, $\blacksquare$ $n=0.37$, $\blacklozenge$ $n=0.27$, $\blacktriangle$ $n=0.20$, $\blacktriangledown$ $n=0.15$, $\circ$ $n=0.12$. Not all densities are shown.}
  \label{values}
  \end{center}
\end{figure} 
 
 In principle, the aim of our  study  is to obtain the dependence of the free energy on all of those parameters (i-v). While the practical  cost  
of the simulations restricts the number of points one can study, we still had generated more than hundred thousand
runs and multiple plots. However, most of it neither can nor should be included in the paper.
Since our physics goal is to understand the back reaction of the dyon ensemble on the holonomy,
we study the whole range of holonomies, from $\nu=0$ to $\nu=1/2$, and only then locate
its minimum. As for the Debye mass, we will find it from the potential and then 
show only the ``selfconsistent" input set.

 What we  actually
need to describe at the end is not the free energy in the whole multi-dimensional space of all parameters,
but the location of the free energy minima. The resulting 
set should be of co-dimension 1, since the original physical setting 
of the problem -- the gauge theory at finite temperature -- has only one input parameter, $T$.

Using the definition of the Debye mass $\frac{g^2}{2V}\frac{\partial ^2 F}{\partial ^2 v} = M_D^2$ for fixed density we get the configurations response to changing the holonomy which is the Debye mass. We require that the used value for the Debye mass is the same as the one found from the derivative of $F$, or atleast not more than $0.4$ below the used value.

The results shows that as the Debye mass goes to zero around the phase transition the only configuration that is consistent with this is that of equal $M$ and $L$ dyons.

\section{The physical results} \label{sec_results} 
We now show only the result 
 which fulfill the self-consistency requirement. Without  fermions the results are symmetric in $\nu \to 1-\nu$ and the results are therefore only for $\nu \leq 1/2$. We have included the Diakonov determinant, though its impact is not too great due to the not so small Debye mass which has been calculated using 3 points. The results here are shown for a wall of $2/(2\pi \nu)$ which was chosen in order to have a large enough density of dyons to overcome the purturbative potential, without completely making the perturbative potential irrelevant. We used $\Lambda =1.5$ to obtain a phase shift around $S=6$. Action is related to temperature as explained in appendix \ref{secHolonomy}. This should of course be fitted to numerical data, but the present data on dyons does not have a high enough efficiency of detection to do this. The action goes up to $S=13$, beyond this value the number of $L$ dyons become too close to 1, and we would need a higher total of dyons to proceed.

\begin{figure}[t!]
  \begin{center}
  \includegraphics[width=7cm]{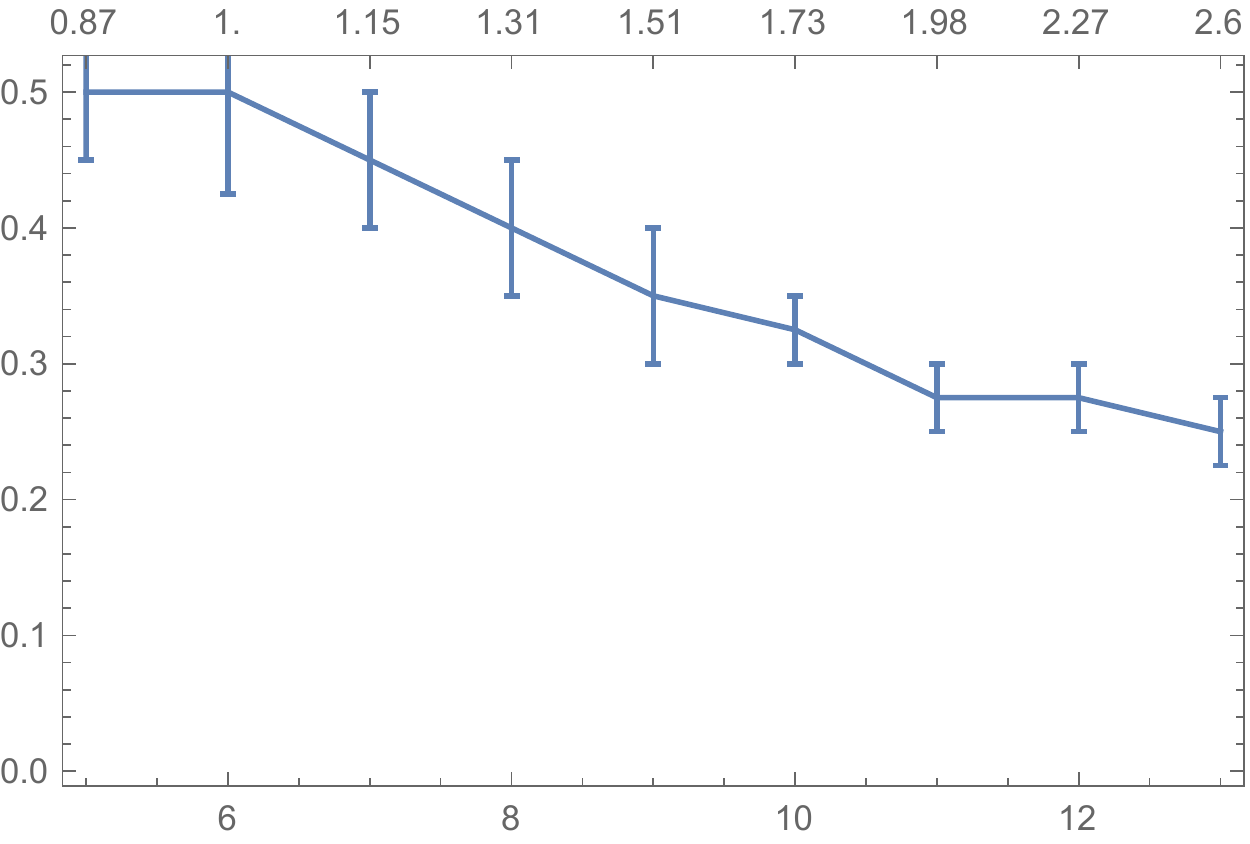}
  \put(-95,-5){$S$}
  \put(-105,140){$T/T_c$}
    \put(-210,60){$\nu$}\\  
    \includegraphics[width=7cm]{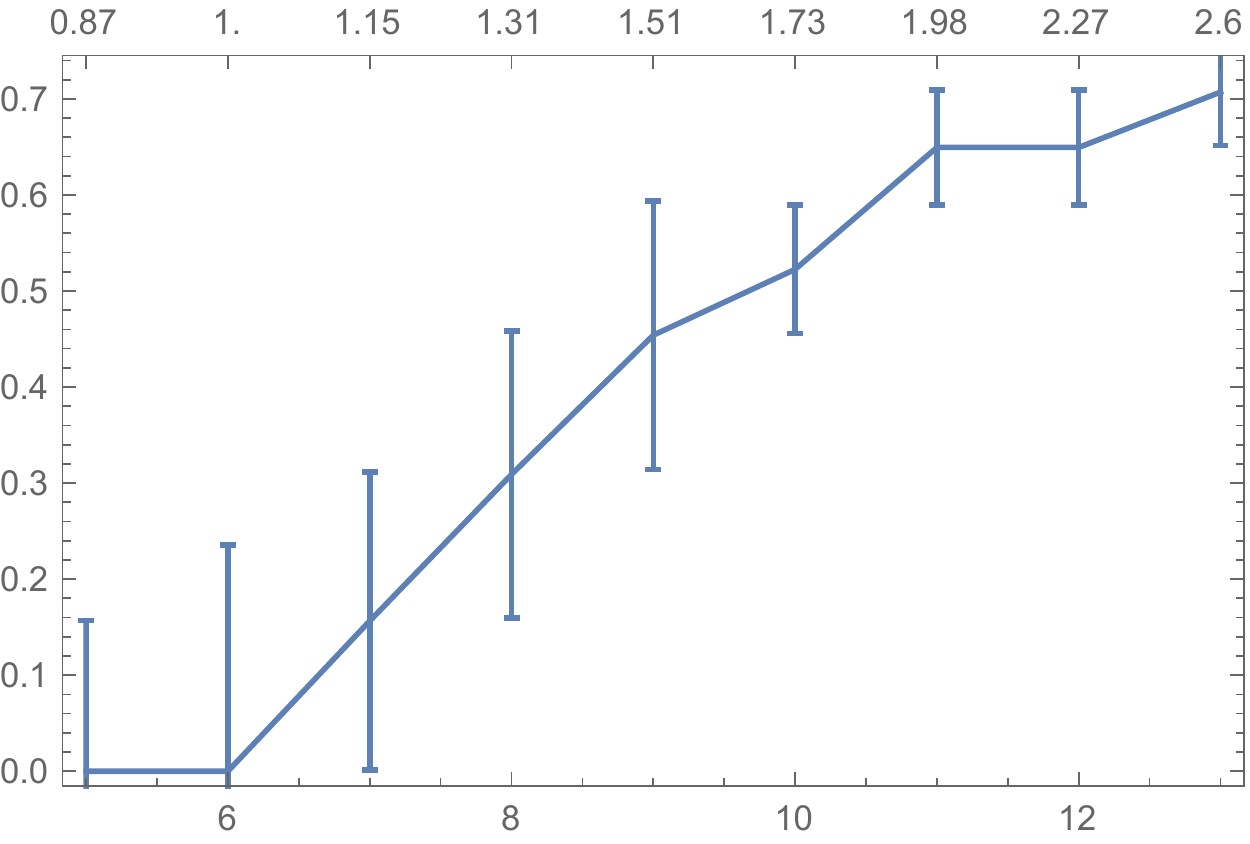}
  \put(-95,-5){$S$}
  \put(-105,140){$T/T_c$}
    \put(-210,60){$P$}
   \caption{  Self-consistent value of the holonomy $\nu$ (upper plot) and Polyakov line (lower plot)
   as a function of action $S$ (lower scales), which is related to $T/T_c$ (upper scales). The error bars are estimates based on the fluctuations of the numerical data.}
  \label{Poly_det}
  \end{center}
\end{figure}

\begin{figure}[t!]
  \begin{center}
  \includegraphics[width=7cm]{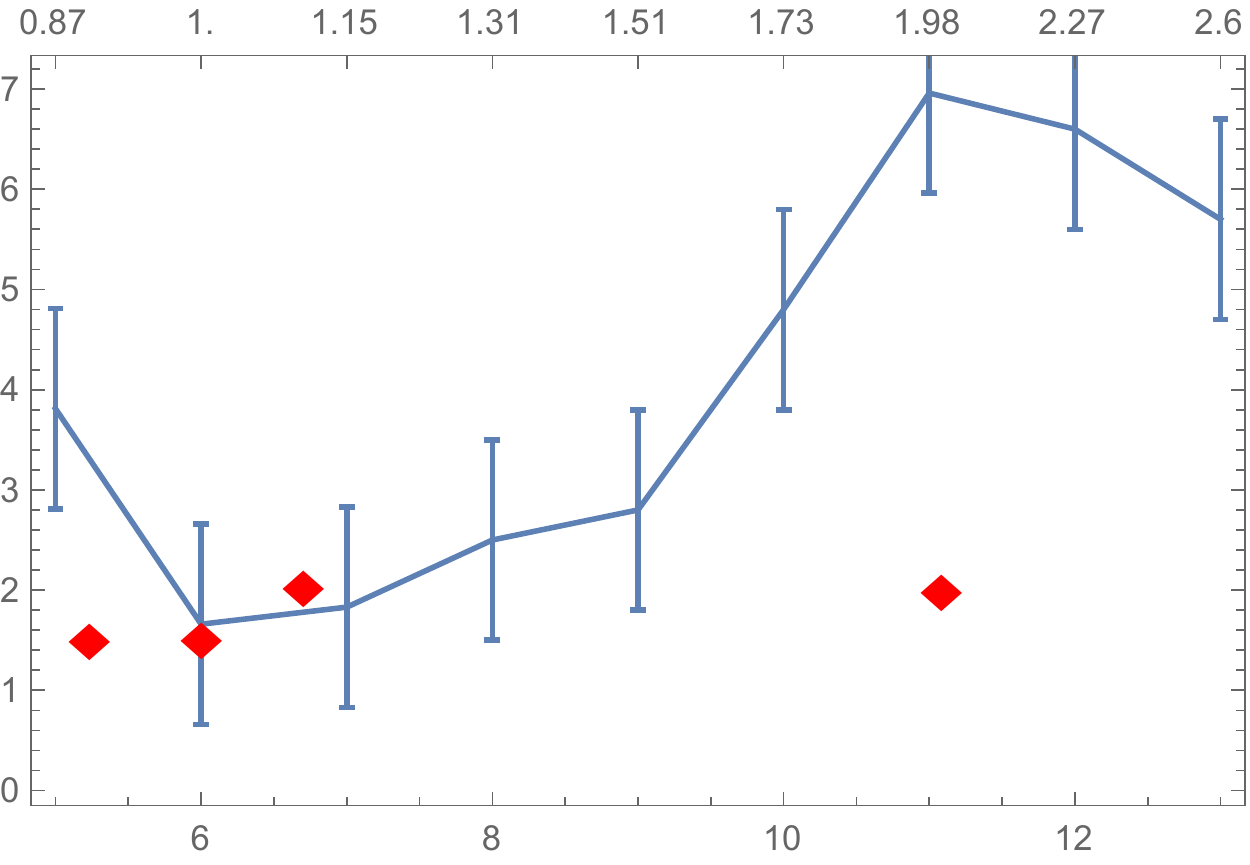}
  \put(-95,-5){$S$}
  \put(-105,140){$T/T_c$}
    \put(-220,60){$M_D$}
   \caption{ (Color online). Self-consistent value of the Debye Mass $M_D$ as a function of action $S$ (lower scale) which is related to $T/T_c$ (upper scale). The error bars are estimates based on the fluctuations of the numerical data. Points represent lattice data from \cite{Bornyakov:2010nc}  as a function of $T/T_c$.}
  \label{Debye_Det}
  \end{center}
\end{figure}

\begin{figure}[t!]
  \begin{center}
  \includegraphics[width=7cm]{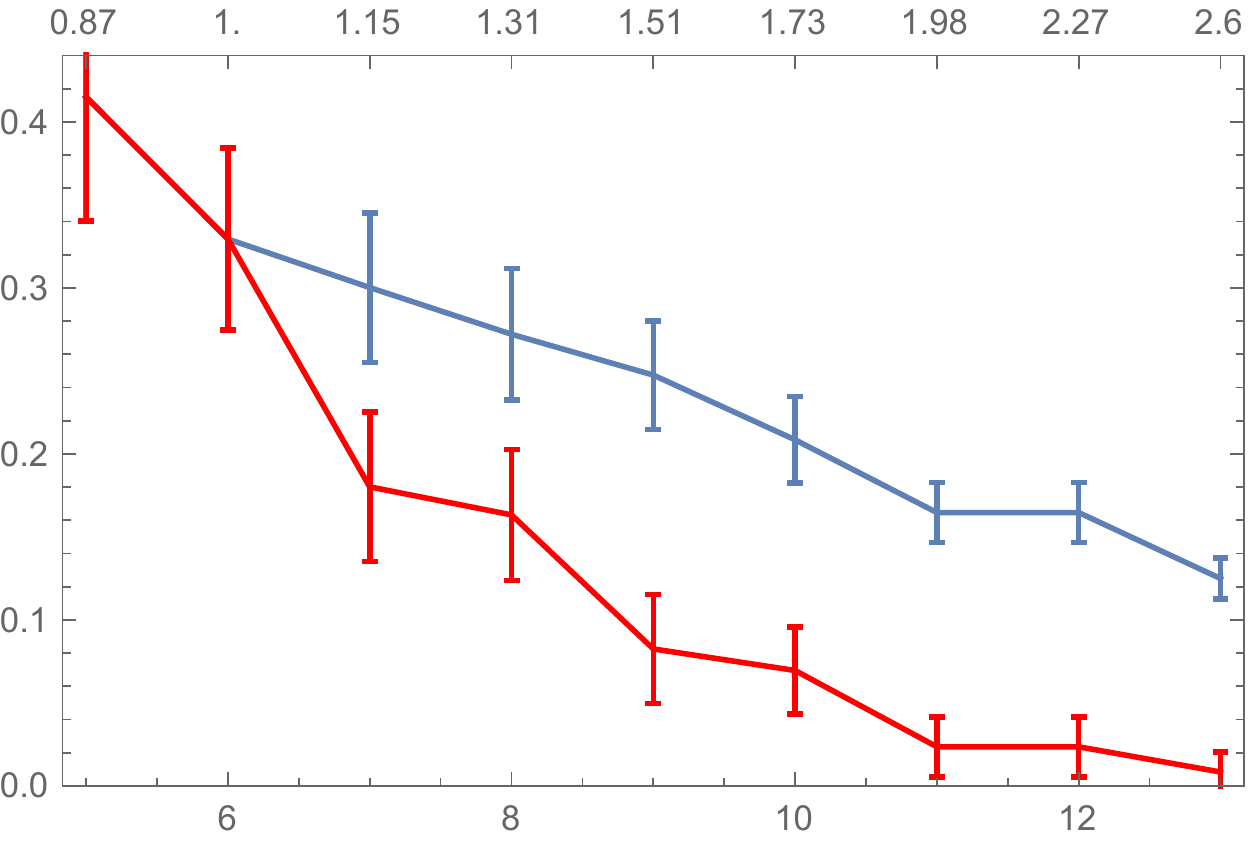}
  \put(-95,-5){$S$}
  \put(-105,140){$T/T_c$}
    \put(-210,60){$n$}
   \caption{ (Color online). Density $n$ (of an individual kind of dyons) as a function of action $S$ (lower scale) which is related to $T/T_c$ (upper scale) for M dyons(higher line) and L dyons (lower line). The error bars are estimates based on the density of points and the fluctuations of the numerical data.}
  \label{Density}
  \end{center}
\end{figure}

The first thing to note about the results is that due to the repulsive Coulomb term between dyons and antidyons of different type, the free energy preferred to have a large Debye mass due to cutting off this repulsion. This meant that when the free energy spectrum as a function of holonomy for a fixed density becomes flat, the small Debye mass created a rise in energy. This resulted in a small jump in holonomy, since the configurations with a holonomy of 0.5 but with slightly higher density than the flat ones, would end up with a smaller free energy. As a result we do not get a completely smooth transition though that is hidden by the size of the errors as seen in Figure \ref{Poly_det} and it also means that at $S=6$ the Debye mass never goes completely to zero, as shown in Figure \ref{Debye_Det}, and the density goes slightly more up also as shown in Fig. \ref{Density}. 

\begin{figure}[t!]
  \begin{center}
  \includegraphics[width=7cm]{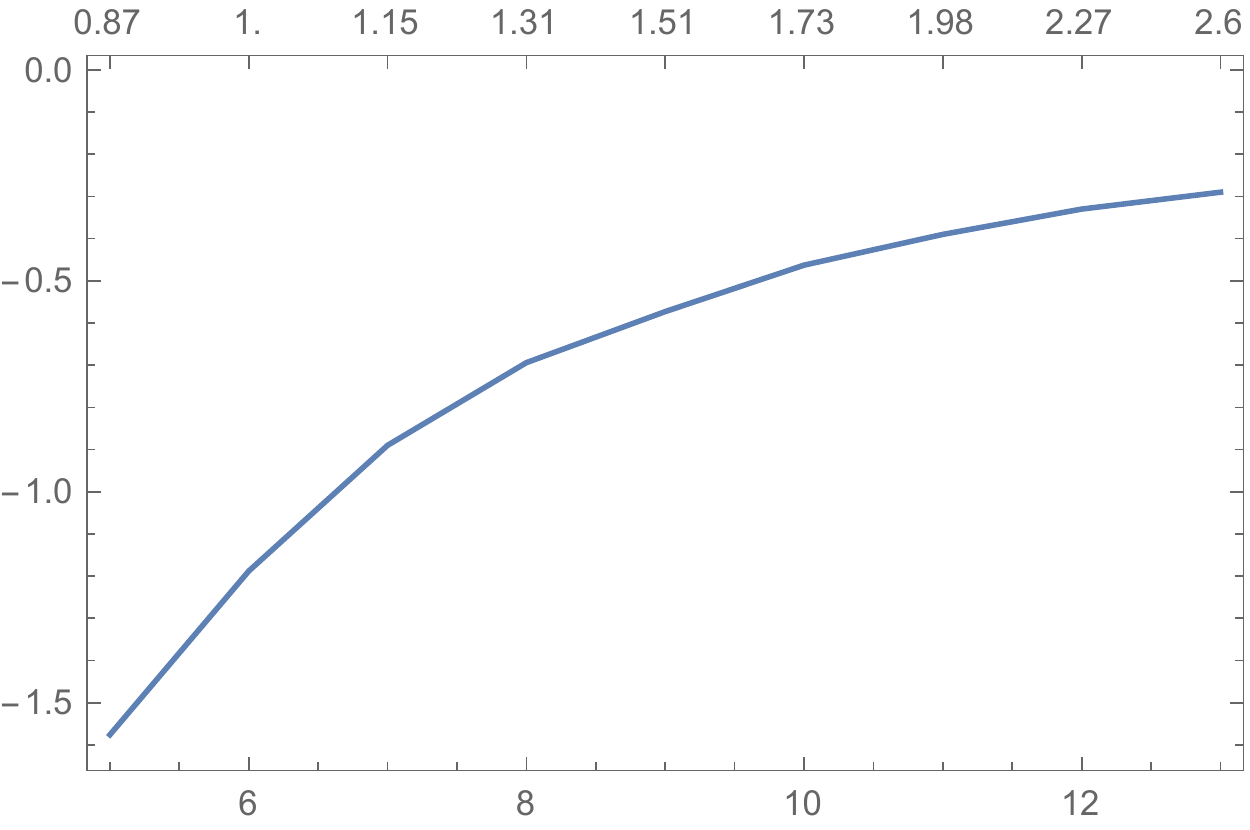}
  	\put(-95,-5){$S$}
  	\put(-105,140){$T/T_c$}
    \put(-210,60){$f$}
   \caption{ (Color online). Self-consistent value of the Free energy density $f$ as a function of action $S$ (lower scale) which is related to $T/T_c$ (upper scale) for $r_0=2$ and $\lambda=1.5$.}
  \label{Md_dM2}
  \end{center}
\end{figure}

\begin{figure}[t!]
  \begin{center}
  \includegraphics[width=7cm]{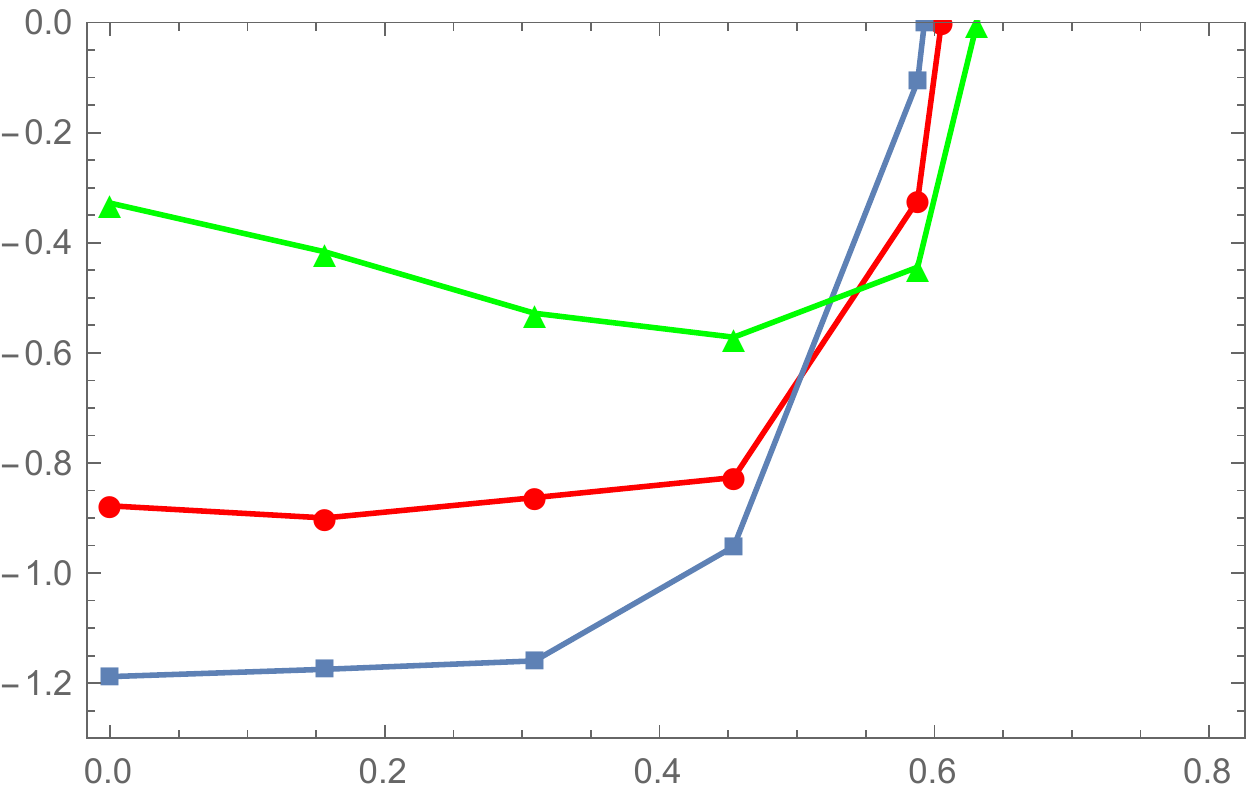}
  \put(-95,-5){$P$}
    \put(-210,60){$f$}
   \caption{ (Color online). (Not-self-consistent in holonomy $\nu$) Free energy density $f$, here shown  as a function of the value of the holonomy
   (in form of the Polyakov loop $P$) at $S=6,7,9$. The lower the action the lower the minimum of the free energy}
  \label{S679}
  \end{center}
\end{figure}

When we are in the confined region we observe the free energy for a fixed density as a single minimum in the middle at $\nu =0.5$. As the action $S$ increases, the density decrease and it becomes more favorable to have some bigger, but lighter dyons, thus shifting the minimum to the sides  as can be seen in Fig. \ref{S679} for $S=6,7,9$. This at the same time makes the lighter dyons, more abundant than the more heavy dyons.

\begin{figure}[t!]
  \begin{center}
  \includegraphics[width=7cm]{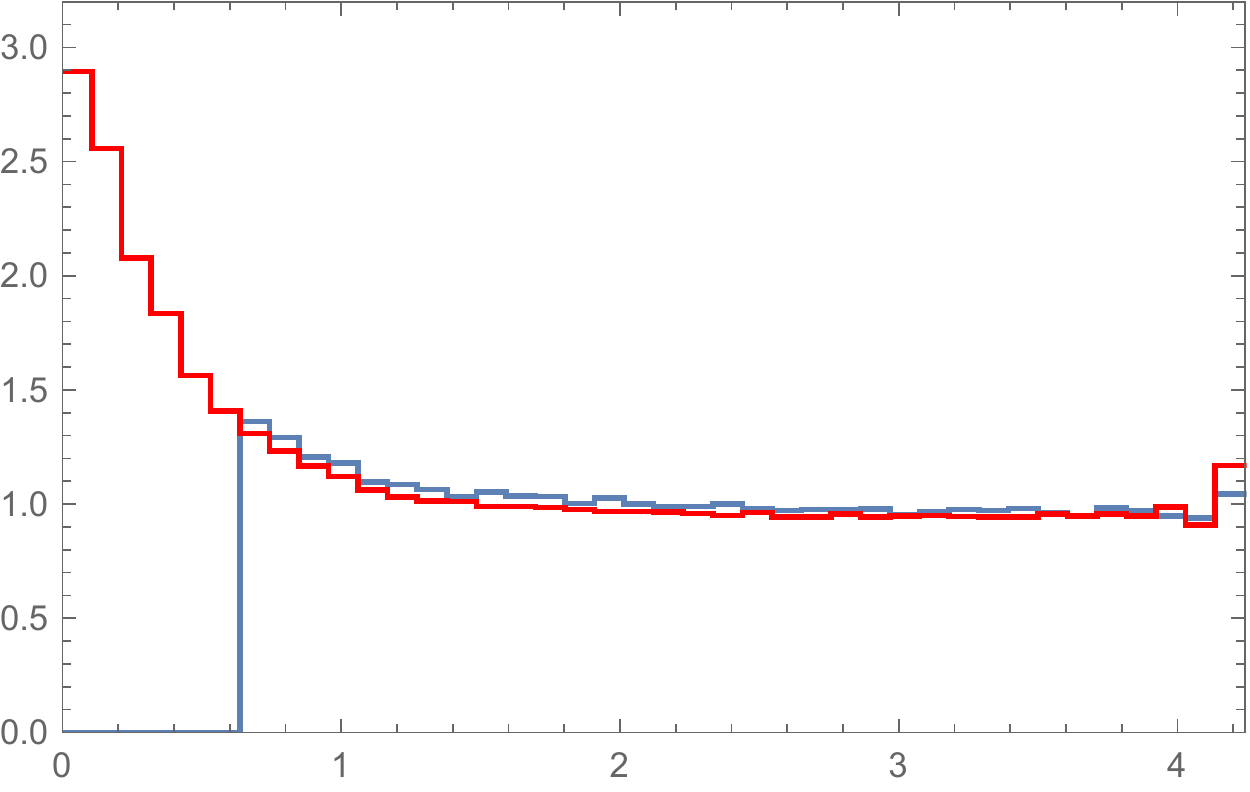}
   \put(-215,60){$C_{ij}$}\\
   \includegraphics[width=7cm]{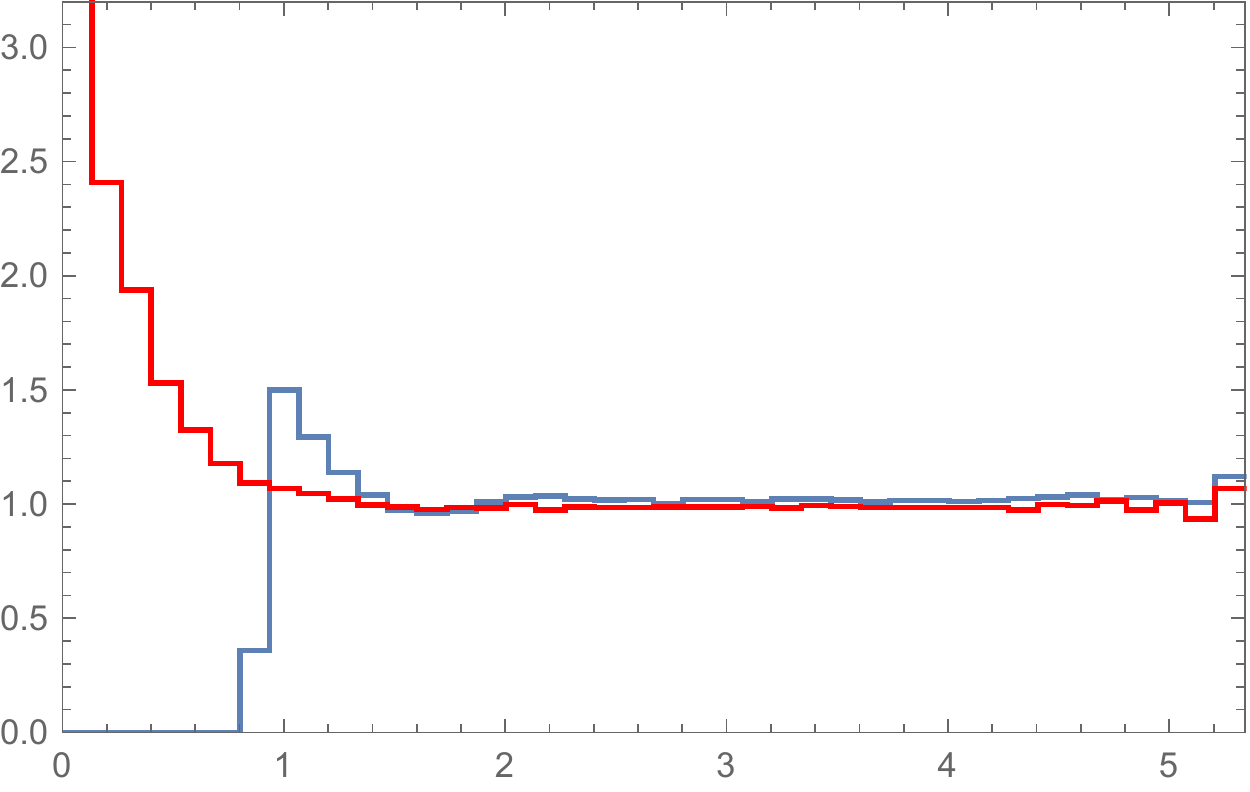}
  \put(-95,-5){$r$}
  \put(-215,60){$C_{ij}$}
   \caption{ (Color online). Correlation function $C_{ij}$ for MM and ML for $S=6$ (upper) and $S=9$ (lower). In the $MM$ case 
 the  correlation function vanishes at small distances due to the core. }
  \label{corre6}
  \end{center}
\end{figure}

Due to the size of the Debye mass, the correlation functions behaves as a liquid with a cutoff at small range. We show the case for $S=6$ in Fig. \ref{corre6} for $MM$ and $ML$.
Note the correlation function  $C_{MM}$  vanishes at small distances due to the core.
The other correlation function $C_{ML}$   for $ML$, displays attraction even at small distances, tripling the density at $r=0$. 
The integrated number of particles in the region in which the correlation function $C_{ML}(r)>1$
 is $0.50$ particles, while for $C_{MM}$ it corresponds to $0.34$ particles: thus the difference is not that large. 

%

\section{Summary and discussion}

As emphasized in the Introduction, an idea that it should be possible to understand confinement (as well as chiral symmetry breaking)   via statistical mechanics in terms of collective coordinates of certain
topological solitons 
goes back to 1970's.  Four decades later we now  have its definite realization
 in terms of the instanton constituent, the instanton-dyons. 
 
 In particular, by identifying classical interaction between  instanton-dyons \cite{Larsen:2014yya}
 and including them in direct Monte-Carlo simulation of the ensemble, together with one-loop
 effects in the measure, we calculated the free energy as a function of all parameters of the model, such as
 the value of the holonomy, dyon densities, and the Debye mass. We then proceed to one-parameter set
 of its minimum, corresponding to dependence on the only left variable, the temperature. The results display
 the deconfinement transition at a  certain density of the dyons. The key to this is the volumes of the dyon 
 repulsive cores, which scale as an inverse cube of the holonomy. 

One of the key questions usually asked in conjecture of such theory is whether 
 the objects we study are sufficiently semiclassical, so let us start our summary with answering  it.
The action per $M$ dyon, $S\nu$, varies in the region studied in this work in the range from 2.5 to 3.3.
Its exponent $exp(-S\nu)$ varies between  0.082 and 0.037. 
The input formula we use include classical and one-loop effects. By selecting specially tuned
$\Lambda$ parameter we basically include the two-loop effects as well. So, we
think that the accuracy of these expressions is sufficient for our purposes.

Since we perform direct simulations using such expressions, we have no further approximations,
and the accuracy of the results is limited only by statistical errors of the Monte Carlo simulations
we make. Let us note that we did not aimed for high  statistical accuracy in this work,
considering it to be a demonstration of the principle. 

What is shown in this work is that the ensemble of instanton-dyons, coupled to holonomy, does undergo
a deconfinement phase transition at certain value of their density. It is physically driven by repulsive interactions,
which enforce ``equality" between $M$ and $L$ dyons, broken in the dilute regime by their
different actions. We see how it happens in detail: first by performing multiple simulations as a
function of all parameters of the model -- dyon densities, holonomy, the value of the Debye mass -- and then
identifying a co-dimension 1 set of the free energy minima, corresponding to physical dyon ensemble as a function of the
temperature $T$.
All these results -- the holonomy potential and the mean Polyakov line $\langle P(T)\rangle$, the dyon densities $n_M(T),n_L(T)$
can and should be compared to lattice data.

This dyon ensemble can be straightforwardly generalized to the QCD-like
theories with arbitrary number of colors and quark flavors. We plan to do
larger scale simulations of those in subsequent publications.

Finally,
let us address a very general question often asked:
why  should one study statistical mechanics of some solitons, rather than directly simulate gauge fields on the lattice, from the first principles? 

Quantum field theories have infinitely many degrees of freedom, and an understanding of which ones
are responsible for a particular phenomenon is very important. 
Using an analogy to condense matter physics: One can
in principle do direct simulations of all electrons in a piece of metal. And yet, understanding the zone
structure, location and shape of the Fermi surfaces offer a much simpler and more intuitive  
approaches to 
 metal thermodynamics and kinetics. To a large extent, the same is true for quarks in
the ``zero mode zone" of the topological solitons. Now we see that  instanton-dyons generate confinement as well
as chiral symmetry breaking.
The model we use have only few variables per $fm^3$ volume, 5-6 orders of magnitude less
than  current lattice simulations. 

\vskip .25cm \textbf{Acknowledgments.} \vskip .2cm 
This work was supported in part by the U.S. Department of Energy under Contract No. DE-FG-88ER40388.

\appendix

\section{Units and holonomy}\label{secHolonomy}
The main physical quantity of the problem is the temperature $T$: it defines the magnitude of the $A_4^3=2\pi\nu T$
(holonomy), the  physical size of the dyons and every other dimensional parameter of the problem.
Yet, precisely because of its omnipresence in the theory at its classical level, dealing only with the dimensionless quantities -- e.g. the
dyon density normalized as $n/T^3$ -- one can in {\em zeroth} approximation cancel all powers of $T$.
At this level, our theory has only dimensionless input parameters.  Most of them --  the dimensionless  
{\em dyon densities}, {\em holonomy} and the {\em the Debye screening mass} -- will be defined
selfconsistently, from the minimum of the free energy. The remaining input will be
 the {\em instanton action parameter} $S$, 
 used in many plots in the text.

   Standard Euclidean formulation of the gauge theory at finite temperature $T$ introduces periodic 
 (Matsubara) time $\tau$ defined on a circle with a  period equal to the inverse temperature $1/T$.
 The exponential of the gauge invariant 
integral over this circle, known as  
  the  Polyakov line
 \be P=   {1\over N_c} Tr Pexp[i \oint A_4^3 (\sigma^3/2) d\tau] \ee
which is gauge invariant due to periodicity.
Here $\sigma ^3$ is the 3rd Pauli matrix.

 As a function of temperature its expectation value
$\langle P\rangle$ changes from 1 at high $T$ to (near) zero at the deconfinement temperature $T_c$. 
In the simplest SU(2) gauge theory we will discuss in this work $\langle P\rangle =cos(\nu \pi)$,
and the holonomy parameter (or just holonomy, for short) $\nu$ changes from 0 to 1/2.
  What remains unknown is the physical origin of this potential.

   Perturbatively, the effect of the holonomy is appearance of nonzero masses of quarks and
   (non-diagonal) gluons, and the corresponding  Gross-Pisarski-Yaffe holonomy potential \cite{Gross:1980br} 
  \be {V_{GPY}(\nu) \over T^4 V_3  } &=& \frac{(2\pi) ^2 \nu ^2\bar{\nu}^2}{3}  \label{GPY}
  \ee
  where $V_3$ is the 3-volume of the box and 
   \be \bar{\nu} &=& 1-\nu \ee
is ``dual holonomy" . 
   We proceed in the text to use dimensionless units for volume $\tilde V_3=T^3 V_3$, densities $n_M=\frac{N_M}{\tilde V_3}$, $n_L=\frac{N_L}{\tilde V_3}$, distances $rT=x$ and free energy density $\frac{ F}{T \tilde V_3 } = \ f$.
Potential $V_{GPY}$
 has a minimum at trivial holonomy $\nu=0$ and a maximum at confining holonomy 
$\nu=1/2$,  thus disfavoring confinement. 

In the $next$ approximation  the so called  quantum  loop effects are incorporated. As is well known,
they lead  to a  running coupling constant. Thus the action parameter (and all others, of course)
become a function of  the basic physical scale given by the temperature $T$.    
 For example, recalling classical instanton action  and the asymptotic freedom formula
 \be S(T)={8\pi^2 \over g^2(T)}=b \cdot ln\left({T\over \Lambda}\right),   \,\, b={11\over 3} N_c \ee
 with the power given by the one-loop beta function.
 If so, the semiclassical factors defining the caloron density now depend on $T$, basically as a  power 
 \be {n_{calorons}(T) \over T^4} \sim e^{-S}\sim \left({\Lambda \over T}\right)^b \ee    
 Since the caloron density has been measured on the lattice at different $T$, one can test this expression 
 against the lattice data. In fact it does work, 
see Fig.1 of Ref.\cite{Shuryak:2013tka},
which confirms that the topological solitons remain semiclassical at the temperatures we discuss.

The next question is the value of the parameter $\Lambda$ in the expression for $S$ above. 
 Note, that our parameter $\Lambda$ is proportional to that in multiple other definitions, such as $\Lambda_{lattice}$
or $\Lambda_{\bar{MS}}$, but
is not equal to them. In principle, the relation between them is known, and the reader may thus ask why
we don't use such relations, obtained from the first principles. The answer is pragmatic: we believe that
the current accuracy of  them raised in high power, e.g. $\Lambda_{\bar{MS}}^b$, is still 
 lower than what was  found from the fit to the caloron data just mentioned. In other words, the  measurements of 
the caloron density is basically the measurements of the high power of $\Lambda$, and they thus provide 
 more accurate values than what can possibly be done  by (much more accurate) measurements
 but of quantities depending on this parameter logarithmically.

Not surprisingly, in practice the meaning (and the value) of $\Lambda$ depends on the context in which
it is used. The fit shown in
 Fig.1 of Ref.\cite{Shuryak:2013tka} corresponds to non-interacting gas of calorons, and it
  give 
\be \Lambda_{calorons} = 0.36 T_c \ee
where $T_c$ is the deconfinement transition temperature, defined in the same lattice work.
If so, the (instanton action) parameter  is $S\approx 7.5$ at $T_c$.  

In our work we worked out much more sophisticated model of the interacting dyon plasma. 
In this model the deconfinement transition happens at a somewhat different value
of the (instanton action) parameter $S\approx 6$. In other words, 
\be \Lambda_{dyonic \, plasma} = 0.44 T_c \ee
 This value is assumed in all plots in section \ref{sec_results} in which our input parameter $S$ is mapped
to the
temperature $T$. 

The reader should however keep in mind, that this mapping between 
the input parameter of the model $S$ and physical $T$ is provisional, it depends
on the model itself.  No doubt it should be
subject to
future improvements, both in
 the lattice data quality used for such fits, or the fit itself. In particular, one should include measurements of the
 instanton-dyons, including the effects of their interaction as discussed in the bulk of our paper.

\begin{table}[h]
\begin{tabular}{ l | c | c | c | r  }
  \hline  
 &  $M$ & $\bar{M}$ & $L$ & $\bar{L}$ \\   
   \hline                     
e & 1 & 1 & -1 & -1 \\
m & 1 & -1 & -1 & 1 \\
h & 1 & 1 & -1 & -1 \\
  \hline  
\end{tabular}
\caption{Quantum numbers of the four different  kinds of the instanton-dyons of the SU(2) gauge theory.
The first two rows are electric and magnetic charges, while by $h$ we mean the contribution
from nonlinear terms including the holonomy field.
}
\label{tab1}
\end{table} 

\section{Instanton-dyons}\label{secDyons}
We do not present here extensive introduction on the configurations and their history, which can be found
in literature such as  \cite{Diakonov:2009ln}. 

``Higgsing" the SU(2) gauge theory by nonzero VEV of $A_4$ called $v$ leads to two massive
and one massless gluons. 
The simplest gauge is the so called regular (hedgehog) gauge, in which the color direction of the  ``Higgs" field is at large r along the unit radial vector  
$A_4^m\rightarrow v \hat r^m$. The solutions are  
\begin{eqnarray}
A_4 ^a &=& \pm \hat{r}_a \left( \frac{1}{r} -v \coth (v r) \right) \nonumber \\
A_i ^a &=& \epsilon _{aij} \hat{r}_j \left(\frac{1}{r}-\frac{v}{\sinh  (v r)}\right), \label{Afield}
\end{eqnarray}
where $+$ corresponds to the $M$ dyon and $-$ corresponds to the $\bar{M}$ dyon. $r$ is the length in position space. The $L$ and $\bar{L}$ dyon are obtained by a replacement $v \to 2 \pi T -v$. To study the classical interaction of the dyons, a gauge transformation is done to make $A_4$ field point in a specific direction (normally this is chosen to be $A_4^3$), which introduced a time dependence in the $L$ dyons in order to to compensate for the extra $2\pi T$. The classical interaction between the dyons can at long range be described by the same formula for all

\be
V(r) &=& \frac{8\pi^2 \nu }{ g^2}\left( (e_1e_2-2h_1h_2)\frac{1}{x}+m_1m_2\frac{1}{x}\right) \\
x &=& 2\pi \nu r \nonumber
\ee
where $e,m,h$ are listed in the Table 1.

As a result sectors that are completely self-dual or anti-self-dual have no interaction, while dyons and antidyons of same type attract and dyons and antidyons of different type repel.

\section{The dyon weights in the partition function}\label{secPartition}
The KvBLL caloron partition function \cite{Diakonov:2004jn} has the form
\be
Z_{KvBLL} &=& \int d^3 z_1 d^3 z_2 T^6 C \left( \frac{8\pi ^2}{g^2} \right) \left( e^{-\frac{8\pi ^2}{g^2}}\right) \left(\frac{1}{T r_{12}}\right)^{\frac{5}{3}} \nonumber \\
  &\times &  ( 2\pi + 4\pi^2 \nu \bar{\nu} T r_{12}) (2 \pi \nu T r_{12}+1)^{\frac{8\nu}{3}-1} \nonumber \\
&\times &  (2 \pi \bar{\nu } T r_{12}+1)^{\frac{8\bar{\nu}}{3}-1} 
  \exp( -V_3T^3 \frac{4 \pi ^2}{3}\nu ^2 \bar{\nu}^2)
\ee
Taking the limit to very dilute situation we find that all powers of $Tr_{12}$ not in the exponential cancel, and we end with
\be
Z_{KvBLL} &=& \int d^3 z_1 d^3 z_2 T^6 C \left( \frac{8\pi ^2}{g^2} \right) \left( e^{-\frac{8\pi ^2}{g^2}}\right)  \nonumber \\
  &\times &  
  (2 \pi \nu )^{\frac{8\nu}{3}}(2 \pi \bar{\nu } )^{\frac{8\bar{\nu}}{3}} 
 \nonumber \\
  &\times & \exp( -V_3T^3 \frac{4 \pi ^2}{3}\nu ^2 \bar{\nu}^2)  \label{B2}
\ee
The term in the exponential corresponds to the perturbative holonomy potential. The Diakonov determinant which we have included is seen to return to a product of the holonomies in the dilute limit
\be
\lim _{T r_{12} \to \infty }\det G &= & \prod _i 4 \pi \nu_i \label{B3}
\ee
By comparison we see that we have to take equation \ref{B2} and divide by equation \ref{B3} in order to get the correct weight for our partition function. We thus end up with the partition function for a $M$ and $L$ dyon given by
\be
Z_{KvBLL} &=& \int d^3 z_1 d^3 z_2 T^6 C \left( \frac{8\pi ^2}{g^2} \right) \left( e^{-\frac{8\pi ^2}{g^2}}\right)  \nonumber \\
 & \times &  \frac{(2 \pi)^{8/3}}{(4\pi)^2}  \nu ^{\frac{8\nu}{3}-1} \bar{\nu } ^{\frac{8\bar{\nu}}{3}-1} \nonumber \\
 &\times &  \exp( -V_3T^3 \frac{4 \pi ^2}{3}\nu ^2 \bar{\nu}^2)
\ee
We redefine the constant so the equation is easier to work with
\be Z_{KvBLL} &=& \int d^3 z_1 d^3 z_2 T^6 \frac{ \Lambda^2}{(4\pi)^2} \left( \frac{8\pi ^2}{g^2} \right)^4 e^{-\frac{8\pi ^2}{g^2}} \nonumber \\
& & \times \nu ^{\frac{8\nu}{3}-1}\bar{\nu} ^{\frac{8\bar{\nu}}{3}-1}
\ee

\end{document}